# Multi-Attention-Network for Semantic Segmentation of Fine Resolution Remote Sensing Images

Rui Li, Shunyi Zheng, Chenxi Duan, Ce Zhang, Jianlin Su, and P.M. Atkinson

**Abstract**—Semantic segmentation of remote sensing images plays an important role in a wide range of applications including land resource management, biosphere monitoring and urban planning. Although the accuracy of semantic segmentation in remote sensing images has been increased significantly by deep convolutional neural networks, several limitations exist in standard models. First, for encoder-decoder architectures such as U-Net, the utilization of multi-scale features causes the underuse of information, where low-level features and high-level features are concatenated directly without any refinement. Second, long-range dependencies of feature maps are insufficiently explored, resulting in sub-optimal feature representations associated with each semantic class. Third, even though the dot-product attention mechanism has been introduced and utilized in semantic segmentation to model long-range dependencies, the large time and space demands of attention impede the actual usage of attention in application scenarios with large-scale input. This paper proposed a Multi-Attention-Network (MANet) to address these issues by extracting contextual dependencies through multiple efficient attention modules. A novel attention mechanism of kernel attention with linear complexity is proposed to alleviate the large computational demand in attention. Based on kernel attention and channel attention, we integrate local feature maps extracted by ResNeXt-101 with their corresponding global dependencies and reweight interdependent channel maps adaptively. Numerical experiments on three large-scale fine resolution remote sensing images captured by different satellite sensors demonstrate the superior performance of the proposed MANet, outperforming the DeepLab V3+, PSPNet, FastFCN, DANet, OCRNet, and other benchmark approaches.

## I. INTRODUCTION

Semantic segmentation of remote sensing images (i.e., the assignment of definite categories to groups of pixels in an image), plays a crucial role in a wide range of applications such as land resources management, yield estimation and economic assessment [1-6].

Vegetation indices are commonly used features extracted from multispectral and hyperspectral images to characterize land surface physical properties. The normalized difference vegetation index (NDVI) [7] and soil-adjusted vegetation index (SAVI) [8] highlight vegetation over other land resources, whereas the normalized difference bareness index (NDBaI) [9] and the normalized difference bare land index (NBLI) [10] emphasize bare land. The normalized difference water index (NDWI) [11] and modified NDWI (MNDWI) [12] indicate water. These indices have been developed and applied widely in the remote sensing community. Meanwhile, different classifiers have been designed from diverse perspectives, from traditional methods such as logistic regression [13], distance measures [14] and clustering [15], to more advanced machine learning methods such as the support vector machine (SVM) [16], random forest (RF) [17] and artificial neural networks (ANN) [18] including the multi-layer perceptron (MLP) [19]. These classifiers depend critically on the quality of features that are extracted for pixel-level land cover classification. However, this high dependency on hand-crafted descriptors restricts the flexibility and adaptability of these traditional methods.

Deep Learning (DL), a powerful approach to capture nonlinear and hierarchical features automatically, has had a significant impact on various domains such as computer vision (CV) [20], natural language processing (NLP) [21] and automatic speech recognition (ASR) [22]. In the field of remote sensing, DL methods have been introduced and implemented for land cover and land use classification. Compared with vegetation indices, which are based on physical and mathematical concepts and hand-coded from spectral bands only, DL methods can mine different kinds of information including temporal periods, spectra, spatial context and the interactions among different land cover categories.

For remotely sensed semantic segmentation, Fully Convolutional Network (FCN)-based methods [5] and encoder-decoder architectures such as SegNet [23] and U-Net [24] have been adopted widely. Generally, the FCN-based architectures comprise a contracting path that extracts information from the input image and generates high-level feature maps, and an

This work was supported in part by the National Natural Science Foundation of China (No. 41671452). *(Corresponding author: Rui Li.)*

R. Li and S. Zheng are with School of Remote Sensing and Information Engineering, Wuhan University, Wuhan 430079, China (e-mail: lironui@whu.edu.cn; syzheng@whu.edu.cn).

C. Duan is with the State Key Laboratory of Information Engineering in Surveying, Mapping, and Remote Sensing, Wuhan University, Wuhan 430079, China; chenxiduan@whu.edu.cn (e-mail: chenxiduan@whu.edu.cn).

C. Zhang and P. M. Atkinson are with the Lancaster Environment Centre, Lancaster University, Lancaster LA1 4YQ, U.K. (e-mail: c.zhang9@lancaster.ac.uk and pma@lancaster.ac.uk).

Jianlin Su is with the Shenzhen Zhuiyi Technology Co., Ltd. (e-mail: bojonesu@wezhuiyi.com).

expanding path, where high-level feature maps are utilized to reconstruct the mask for pixel-wise segmentation by the single [5] or multi-level [24, 25] up-sampling procedures. Despite their powerful representation capability, however, information flow bottlenecks limit the potential of these multi-scale approaches [26]. For example, the low-level and fine-grained detailed feature maps generated by the encoder are concatenated with high-level and coarse-grained semantic information generated by the decoder without any further refinement, leading to inadequate exploitation and deficient discrimination of features. Besides, the discriminative ability of the feature representations might be insufficient for challenging tasks such as semantic segmentation of fine spatial resolution remote sensing images.

The utilization of context fusion at multiple scales is a feasible solution [27-33], increasing the discriminative power of feature representations. The multi-scale contextual information can be aggregated using techniques such as atrous spatial pyramid pooling [27, 28], pyramid pooling module [29], or context encoding module [31]. Although context captured by the above strategies is beneficial to characterizing objects at different scales, the contextual dependencies for whole input regions are homogeneous and non-adaptive, without considering the disparity between contextual dependencies and local representation of different categories. Further, these multi-scale context fusion strategies are designed manually, with limited flexibility in modeling multi-context representations. The long-range dependencies of feature maps are insufficiently leveraged in these approaches, which may be of paramount importance for remotely sensed semantic segmentation.

With strong capabilities to capture long-range dependencies, dot-product attention mechanisms have been applied in vision and natural language processing tasks. The dot-product-attention-based Transformer has demonstrated state-of-the-art performance in a majority of tasks in natural language processing [21, 34-36]. The non-local module [37], a dot-product-based attention modified for computer vision, has shown great potential in image classification [38], object detection [39], semantic segmentation [40] and panoptic segmentation [41].

Utilization of the dot-product attention mechanism often comes with significant memory and computational costs, which increase quadratically with the size of the input over space and time. It remains an intractable problem to model global dependency on large-scale inputs, such as video, long sequences and high-resolution images. To alleviate the substantial computational requirement, Child et al. [42] designed a sparse factorization of the attention matrix and reduced the complexity from $O(N^2)$ to $O(N\sqrt{N})$. Using locality sensitive hashing, Kitaev et al. [43] reduced the complexity to $O(N \log N)$. Katharopoulos et al. [44] represented self-attention as a linear dot-product of kernel feature maps to further reduce the complexity to $O(N)$, and Shen et al. [45] modified the position of the softmax functions.

In this paper, by comparison, we not only dramatically decrease the complexity, but also amply exploit the potential of the attention mechanism by designing a multilevel framework.

Specifically, we reduce the complexity of the dot-product attention mechanism to $O(N)$ by treating attention as a kernel function. As the complexity of attention is reduced dramatically by kernel attention, we propose a Multi-Attention-Network (MANet) with a ResNeXt-101 [46] backbone which explores the complex combinations between attention mechanisms and deep networks for the task of semantic segmentation using fine-resolution remote sensing images. The performance of the proposed algorithm is compared comprehensively with benchmarks U-Net [24], DeepLab V3 [27], DeepLab V3+ [28], RefineNet [25], PSPNet [29], FastFCN [32], DANet [40] and OCRNet [47]. The major contributions of this research include:

1) A novel attention mechanism involving kernel attention with linear complexity is proposed to alleviate the attention module's huge computational demand.

2) To extract refined dense features, we replace ResNet with ResNeXt-101 as the backbone, enhancing the ability for feature extraction.

3) Based on kernel attention and ResNeXt-101, we propose a Multi-Attention-Network (MANet) by extracting contextual dependencies using multi-kernel attention.

## II. RELATED WORK

### A. Attention Inspired by Human Perception

Due to the overwhelming computational requirement for perceiving surrounding scenes with detail equivalent to foveal vision, the selective visual attention endows humans with the ability to orientate rapidly towards salient objects in a sophisticated visual scene [48] and choose a subset of the available perceptual information before further processing. Inspired by the human attention mechanism, substantial algorithms have been developed over the last few decades [49-51].

Recently, a very large number of domains has been influenced significantly by the wave of DL, which emphasizes end-to-end hierarchical feature extraction in an automatic fashion. Integration of DL with the attention mechanism has great potential to transform the paradigm in this field. Attention in DL could be regarded as a weighted combination of the input feature maps, where the weights are hinged on the similarities between elements of the input [52]. Given that kernel learning [53] processes all inputs simultaneously and order-independently by computing the similarity between the inputs, attention could be interpreted as a kernel smoother [54] applied over the inputs in a sequence, where the kernel evaluates the similarity between different inputs. The formulae and mathematical proofs can be found in [52].

### B. Dot-Product Attention

To enhance word alignment in machine translation, Bahdanau et al. [55] proposed the initial formulation of the dot-product attention mechanism. Subsequently, recurrences are entirely replaced by attention in the Transformer [56]. State-of-the-art records in most natural language processing tasks demonstrate the superiority of attention mechanisms amongst others. Wang et al. [37] modified dot-product attention for computer vision and proposed the non-local module. This method has been developed and applied to many tasks of computer vision, including image classification [38], object detection [39], semantic segmentation [40] and panoptic segmentation [41]. These successful applications demonstrated further the effectiveness and general utility of attention mechanisms. Moreover, attention also attained state-of-the-art performance in speech recognition [56, 57].

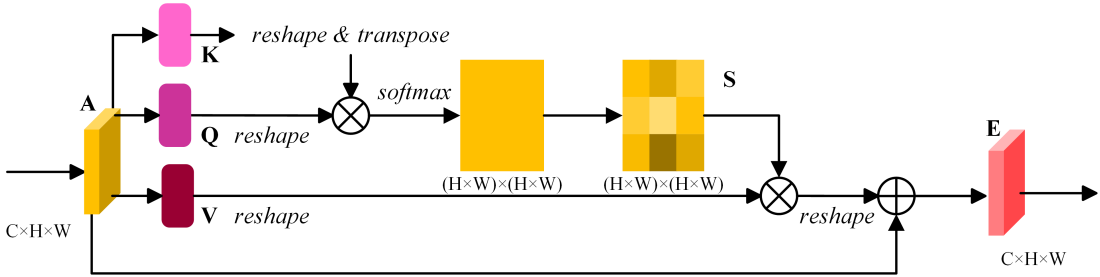

Fig. 1. Illustration of the architecture of the dot-product attention mechanism.

### C. Scaling Attention

Besides dot-product attention, there exists another set of techniques for scaling attention (or simply attention) in the literature. Unlike dot-product attention which models global dependency, scaling attention reinforces informative features and whittles information-lacking features. In the squeeze-and-excitation (SE) module [58], a global average pooling layer and a linear layer are harnessed to calculate a scaling factor for each channel, and then the channels are weighted accordingly. The convolutional block attention module (CBAM) [59], selective kernel unit (SK unit) [60] and efficient channel attention module (ECA) [61] further boost the SE block's performance. The principles and purposes of dot-product attention and scaling attention are entirely divergent. This paper focuses on dot-product attention due to its superiority in many computer vision and pattern recognition tasks.

### D. Semantic Segmentation

FCN-based methods have brought tremendous progress and evolution in semantic segmentation. DilatedFCN and EncoderDecoder are two prominent directions followed by FCN. In DilatedFCNs [27-32, 62], dilate or atrous convolutions are harnessed to retain the receptive field-of-view, and a multi-scale context module is utilized to cope with high-level feature maps. Alternatively, EncoderDecoders [24, 25, 63-69] utilize an encoder to capture multi-level feature maps, which are then incorporated into the final prediction using a decoder.

**DilatedFCN** The dilated or atrous convolution [30, 62] has been demonstrated to be an effective technology for dense prediction and has achieved high accuracy in semantic segmentation. In DeepLab [27, 28], the atrous spatial pyramid pooling (ASPP), comprised of parallel dilated convolutions with diverse dilated rates, is able to embed contextual information, while the pyramid pooling module (PPM) enables PSPNet [29] to incorporate the contextual prior among different scales. Alternatively, EncNet [31] utilizes a context encoding module to exploit global contextual information. FastFCN [32] further replaces the dilated convolutions with a joint pyramid upsampling (JPU) module to reduce computational complexity. To extract abundant contextual relationships, a dot-product attention mechanism is attached to the DANet [40]. For further differentiating the same-object-class contextual pixels from the different-object-class contextual pixels, the object-contextual representation (OCR) module is elaborated by the OCRNet [47].

**EncoderDecoder** Skip connections are employed to integrate the high-level features generated by the decoder and the low-level features generated by the corresponding encoder, which are the essential structure of U-Net [24]. In the recent literature [63-65], the plain skip connections in U-Net are substituted by more subtle and elaborate skip connections which reduce the semantic gap between the encoder and decoder. Meanwhile, the structural development based on residual connections is also a promising direction [25, 66-69]. Taking DeepLab V3 as the encoder, DeepLab V3+ [28] combined the merits of DilatedFCN and EncoderDecoder in a single framework.

### E. Attention-based Networks for Semantic Segmentation

Based on dot-product attention as well as its variants, various attention-based networks have been proposed to cope with the semantic segmentation task. Inspired by the non-local module [37], the Double Attention Networks ($A^2$-Net) [70], Dual Attention Network (DANet) [40], Point-wise Spatial Attention Network (PSANet) [71], Object Context Network (OCNet) [72], and Co-occurrent Feature Network (CFNet) [73] were proposed for scene segmentation by exploring the long-range dependency.

The computing resource required by dot-product attention modules is normally huge, which severely limits the application of attention mechanisms. Therefore, substantial researches have been implemented which aim to alleviate the bottleneck to efficiency and push the boundaries of attention, including accelerating the generation process of the attention matrix [47, 74-76], pruning the structure of the attention block [77], and optimizing attention based on low-rank reconstruction [78].

Meanwhile, another burgeoning research area for semantic segmentation is how to embed the dot-product attention into a Graph Convolutional Network (GCN) and optimize the complexity of the attention [79-83].

### III. METHODOLOGY

#### A. Definition of Dot-Product Attention

Supposing $N$ and $C$ denote the length of input sequences and the number of input channels, respectively, where $N = H \times W$, and $H$ and $W$ denote the height and width of the input, given a feature $X = [x_1, \cdots, x_N] \in \mathbb{R}^{N \times C}$, dot-product attention utilizes three projected matrices $W_q \in \mathbb{R}^{D_x \times D_k}$, $W_k \in \mathbb{R}^{D_x \times D_k}$, and $W_v \in \mathbb{R}^{D_x \times D_v}$ to generate the corresponding query matrix $Q$, key matrix $K$ and value matrix $V$ as:

$$\boldsymbol{Q} = \boldsymbol{X}\boldsymbol{W}_q \in \mathbb{R}^{N \times D_k},$$

$$\boldsymbol{K} = \boldsymbol{X}\boldsymbol{W}_k \in \mathbb{R}^{N \times D_k}, \tag{1}$$

$$\boldsymbol{V} = \boldsymbol{X}\boldsymbol{W}_v \in \mathbb{R}^{N \times D_v}.$$

where $D_{(\cdot)}$ means the dimension of $(\cdot)$. Please note that the shapes of $\boldsymbol{Q}$ and $\boldsymbol{K}$ are supposed to be identical. Therefore, we use the same symbol to represent their shapes.

A normalization function $\rho$ evaluates the similarity between the $i$-th query feature $\boldsymbol{q}_i^T \in \mathbb{R}^{D_k}$ and the $j$-th key feature $\boldsymbol{k}_j \in \mathbb{R}^{D_k}$ by $\rho(\boldsymbol{q}_i^T \boldsymbol{k}_j) \in \mathbb{R}^1$. Please note that the vectors in this paper default to column vectors. Generally, as the query feature and key feature are generated by diverse layers, the similarities between $\rho(\boldsymbol{q}_i^T \boldsymbol{k}_j)$ and $\rho(\boldsymbol{q}_j^T \boldsymbol{k}_i)$ are not symmetric. By calculating the similarities between all pairs of positions and taking the similarities as weights, the dot-product attention module computes the value at position $i$ by aggregating the value features from all positions based on weighted summation:

$$D(\boldsymbol{Q}, \boldsymbol{K}, \boldsymbol{V}) = \rho(\boldsymbol{Q}\boldsymbol{K}^T)\boldsymbol{V}. \tag{2}$$

The softmax is a standard normalization function as:

$$\rho(\boldsymbol{Q}^T\boldsymbol{K}) = \text{softmax}_{\text{row}}(\boldsymbol{Q}\boldsymbol{K}^T), \tag{3}$$

where $\text{softmax}_{\text{row}}$ indicates the application of the softmax function along each row of the matrix $\boldsymbol{Q}\boldsymbol{K}^T$.

The $\rho(\boldsymbol{Q}\boldsymbol{K}^T)$ models the similarities between all pairs of positions. However, as $\boldsymbol{Q} \in \mathbb{R}^{N \times D_k}$ and $\boldsymbol{K}^T \in \mathbb{R}^{D_k \times N}$, the product between $\boldsymbol{Q}$ and $\boldsymbol{K}^T$ belongs to $\mathbb{R}^{N \times N}$, leading to $O(N^2)$ memory complexity and $O(N^2)$ computational complexity. As a consequence, the high resource-demand of the dot-product critically limits its application to large-scale inputs. One way to solve this problem is to modify the softmax [45], and another is to rethink the attention via the lens of the kernel. An illustration of the architecture for the dot-product attention mechanism is shown in Fig. 1.

### B. Generalization of Dot-Product Attention Based on Kernel

Under the condition of the softmax normalization function, the $i$-th row of the result matrix generated by the dot-product attention module (equation 2) can be written as:

$$D(\boldsymbol{Q}, \boldsymbol{K}, \boldsymbol{V})_i = \frac{\sum_{j=1}^N e^{\boldsymbol{q}_i^T \boldsymbol{k}_j} \boldsymbol{v}_j}{\sum_{j=1}^N e^{\boldsymbol{q}_i^T \boldsymbol{k}_j}}, \tag{4}$$

Then, equation (4) can be generalized for any normalization function and rewritten as:

$$D(\boldsymbol{Q}, \boldsymbol{K}, \boldsymbol{V})_i = \frac{\sum_{j=1}^N \text{sim}(\boldsymbol{q}_i, \boldsymbol{k}_j) \boldsymbol{v}_j}{\sum_{j=1}^N \text{sim}(\boldsymbol{q}_i, \boldsymbol{k}_j)}, \tag{5}$$

$$\text{sim}(\boldsymbol{q}_i, \boldsymbol{k}_j) \geq 0.$$

where $\text{sim}(\boldsymbol{q}_i, \boldsymbol{k}_j)$ indicates the function calculating the similarity between $\boldsymbol{q}_i$ and $\boldsymbol{k}_j$. If $\text{sim}(\boldsymbol{q}_i, \boldsymbol{k}_j) = e^{\boldsymbol{q}_i^T \boldsymbol{k}_j}$, equation (5) is equivalent to equation (4). And $\text{sim}(\boldsymbol{q}_i, \boldsymbol{k}_j)$ can be further expanded as $\text{sim}(\boldsymbol{q}_i, \boldsymbol{k}_j) = \phi(\boldsymbol{q}_i)^T \varphi(\boldsymbol{k}_j)$, where $\phi(\cdot)$ and $\varphi(\cdot)$ can be considered as kernel smoothers [52] if $\phi = \varphi$. Accordingly, the corresponding inner product space can be defined as $\langle \phi(\boldsymbol{q}_i), \varphi(\boldsymbol{k}_j) \rangle$.

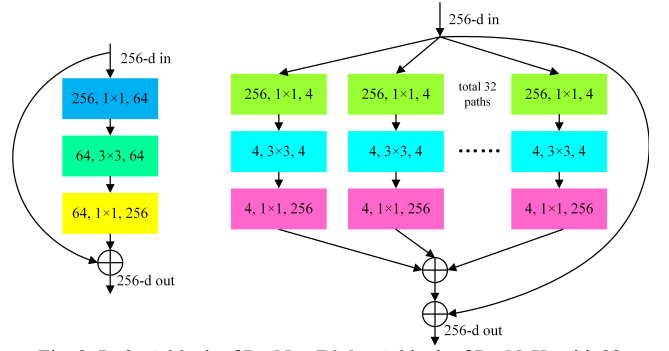

Fig. 2. **Left**: A block of ResNet. **Right**: A block of ResNeXt with 32 cardinalities. Each layer is shown as: in channels, kernel size, out channels.

Equation (4) can then be further rewritten as:

$$D(\boldsymbol{Q}, \boldsymbol{K}, \boldsymbol{V})_i = \frac{\sum_{j=1}^N \phi(\boldsymbol{q}_i)^T \varphi(\boldsymbol{k}_j) \boldsymbol{v}_j}{\sum_{j=1}^N \phi(\boldsymbol{q}_i)^T \varphi(\boldsymbol{k}_j)}, \tag{6}$$

which can be simplified as:

$$D(\boldsymbol{Q}, \boldsymbol{K}, \boldsymbol{V})_i = \frac{\phi(\boldsymbol{q}_i)^T \sum_{j=1}^N \varphi(\boldsymbol{k}_j) \boldsymbol{v}_j^T}{\phi(\boldsymbol{q}_i)^T \sum_{j=1}^N \varphi(\boldsymbol{k}_j)}. \tag{7}$$

As $\boldsymbol{K} \in \mathbb{R}^{D_k \times N}$ and $\boldsymbol{V}^T \in \mathbb{R}^{N \times D_v}$, the product between $\boldsymbol{K}$ and $\boldsymbol{V}^T$ belongs to $\mathbb{R}^{D_k \times D_v}$, which reduces the complexity of the dot-product attention mechanism considerably.

### C. Kernel Attention

We take $\phi(\cdot) = \varphi(\cdot) = \text{softplus}(\cdot)$, where

$$\text{softplus}(x) = \log(1 + e^x). \tag{8}$$

The reason why we select $\text{softplus}(\cdot)$ instead of $\text{ReLU}(\cdot)$ is that the nonzero property of the softplus enables the attention to avoid zero gradients when the input is negative. Then, the similarity function can be embodied as:

$$\text{sim}(\boldsymbol{q}_i, \boldsymbol{k}_j) = \text{softplus}(\boldsymbol{q}_i)^T \text{softplus}(\boldsymbol{k}_j), \tag{9}$$

thereby rewriting the equation (5) as:

$$D(\boldsymbol{Q}, \boldsymbol{K}, \boldsymbol{V})_i = \frac{\text{softplus}(\boldsymbol{q}_i)^T \sum_{j=1}^N \text{softplus}(\boldsymbol{k}_j) \boldsymbol{v}_j^T}{\text{softplus}(\boldsymbol{q}_i)^T \sum_{j=1}^N \text{softplus}(\boldsymbol{k}_j)}, \tag{10}$$

which can be further written in a vectorized form as:

$$D(\boldsymbol{Q}, \boldsymbol{K}, \boldsymbol{V}) = \frac{\text{softplus}(\boldsymbol{Q}) \text{softplus}(\boldsymbol{K})^T \boldsymbol{V}}{\text{softplus}(\boldsymbol{Q}) \sum_j \text{softplus}(\boldsymbol{K})_{i,j}^T}. \tag{11}$$

As $\sum_{j=1}^N \text{softplus}(\boldsymbol{k}_j) \boldsymbol{v}_j^T$ and $\sum_{j=1}^N \text{softplus}(\boldsymbol{k}_j)$ can be calculated and reused for each query, the time and memory complexity of the proposed linear attention mechanism based on equation (11) is $O(N)$ only.

### D. The Structure of the ResNeXt

We utilize the ResNeXt instead of the ResNet as the backbone of proposed MANet. Inspired by ResNet [84] and Inception [85], the ResNeXt [46] structure utilized both the strategy of repeating layers and the technique of stacking and split-convert-merge to increase extensibility and accuracy without increasing the complexity of the network. ResNeXt is constructed by repeating a basic block that aggregates a series of transformations using the same topology. The number of transformations is defined as its cardinality. A comparison between ResNet and ResNeXt is demonstrated in Fig. 2.

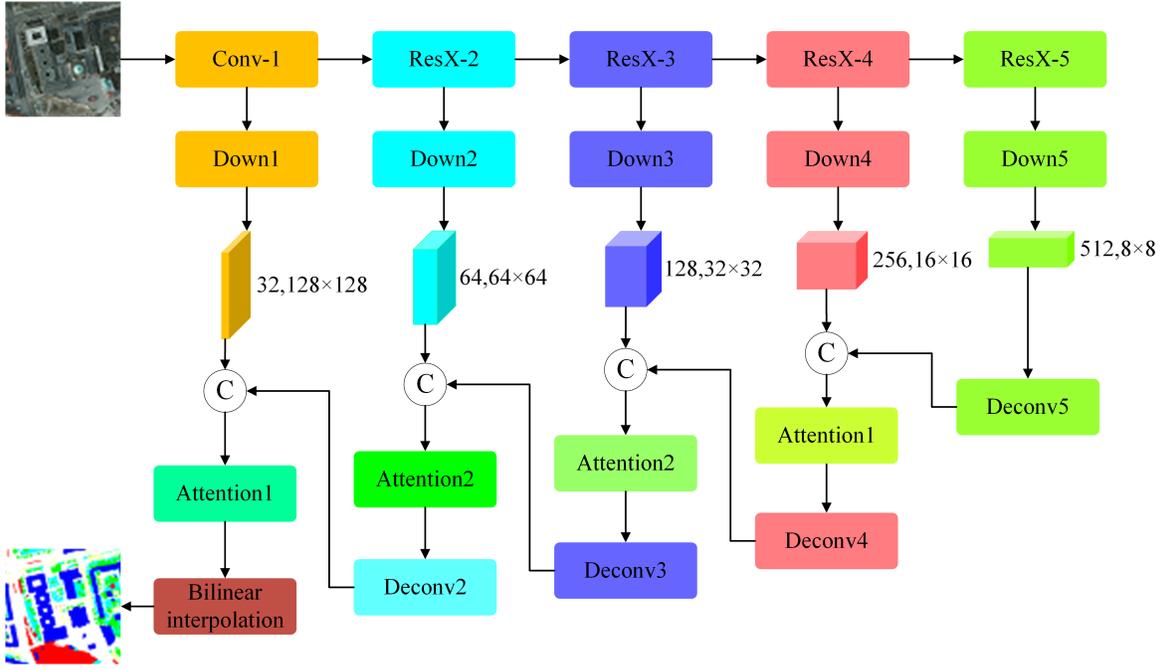

Fig. 5. The structure of the proposed MANet.

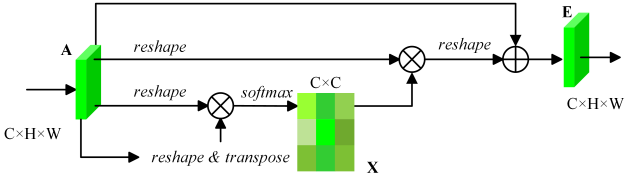

Fig. 3. Details of the channel attention mechanism.

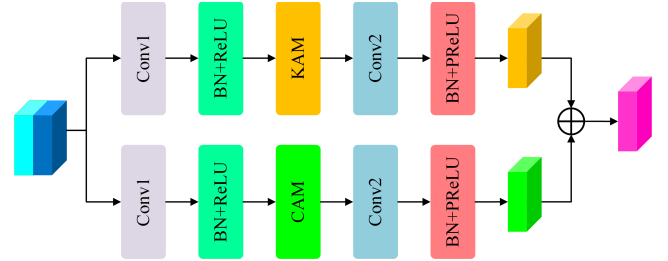

Fig. 4. Illustration of the attention block.

Deepening or widening a network is the conventional approach to increasing the accuracy of the network. Nevertheless, the difficulties of network design and computational complexity increase with the number of hyper-parameters, such as channels and kernel sizes. Alternatively, ResNeXt combines the block stack technology adopted by ResNet and the grouping convolution used in Inception to design an aggregated transformation strategy, with increased accuracy in discrimination without increasing the complexity of the model.

Experiments indicate that the increase in cardinality is a concrete and measurable way to enhance the representation capability of a network [46], especially when deepening or widening a network starts to produce diminishing marginal returns for existing deep networks.

### E. Multi-Attention-Network

For the spatial dimension, as the computational complexity of the dot-product attention mechanism exhibits a quadratic relationship with the size of the input ($N = H \times W$), we design an attention mechanism based on kernel attention, named KAM. For the channel dimension, the number of input channels $C$ is normally far less than the number of pixels contained in the feature maps (i.e., $C \ll N$). Therefore, the complexity of the softmax function for channels (i.e., $O(C^2)$), is not large according to equation (3). Thus, we utilize the channel attention mechanism based on the dot-product [40], named CAM (Fig.

3). Using the kernel attention mechanism (KAM) and channel attention mechanism (CAM) which model the long-range dependencies of positions and channels, respectively, we design an attention block to enhance the discriminative ability of feature maps extracted by each layer (Fig. 4).

The structure of the proposed Multi-Attention-Network is illustrated in Fig. 5. We harness ResNeXt-101 pretrained on ImageNet to extract feature maps. Specifically, five feature maps at different scales acquired from the outputs of [Conv-1, Res-2, Res-3, Res-4, Res-5] are employed. The low-level feature maps are then up-sampled by transposed convolution and concatenated with the high-level feature maps. The concatenated results are subsequently fed into an attention block. Finally, the output of the final attention block is up-sampled to the identical spatial resolution of the input by employing a bilinear interpolation approach.

## IV. EXPERIMENTAL RESULTS AND ANALYSIS

### A. Datasets

The effectiveness of the linear attention mechanism is tested using the ISPRS Potsdam dataset, the ISPRS Vaihingen dataset

TABLE I
THE EXPERIMENTAL RESULTS ON THE ISPRS POTSDAM (THE LEFT) AND VAIHINGEN (THE RIGHT) DATASETS.

| Method | PA | MPA | K | mIoU | FWIoU | F1 | Method | PA | MPA | K | mIoU | FWIoU | F1 |
|---|---|---|---|---|---|---|---|---|---|---|---|---|---|
| U-Net | 83.415 | 82.729 | 77.764 | 71.750 | 71.962 | 83.208 | U-Net | 84.031 | 80.010 | 78.805 | 68.684 | 72.725 | 81.031 |
| Deeplab V3 | 87.661 | 86.207 | 83.483 | 76.825 | 78.343 | 86.702 | Deeplab V3 | 85.557 | 82.767 | 80.822 | 71.918 | 75.052 | 83.377 |
| Deeplab V3+ | 87.812 | 87.180 | 83.640 | 77.568 | 78.556 | 87.125 | Deeplab V3+ | 85.925 | 82.446 | 81.321 | 72.292 | 75.676 | 83.582 |
| RefineNet | 87.079 | 86.282 | 82.720 | 77.011 | 77.644 | 86.745 | RefineNet | 85.433 | 82.333 | 80.654 | 71.904 | 74.856 | 83.346 |
| PSPNet | 87.184 | 86.264 | 82.859 | 76.531 | 77.687 | 86.507 | PSPNet | 86.372 | 82.580 | 81.901 | 72.445 | 76.296 | 83.641 |
| FastFCN | 87.834 | 86.528 | 83.620 | 77.816 | 78.660 | 87.310 | FastFCN | 86.032 | 82.222 | 81.456 | 71.645 | 75.799 | 83.045 |
| DANet | 88.512 | 87.251 | 84.618 | 78.608 | 79.665 | 87.846 | DANet | 86.010 | 82.899 | 81.429 | 72.258 | 75.767 | 83.560 |
| OCRNet | 87.898 | 87.085 | 83.783 | 78.179 | 78.806 | 87.520 | OCRNet | 86.157 | 82.991 | 81.628 | 72.639 | 75.948 | 83.852 |
| MANet | 89.396 | 88.763 | 85.808 | 80.585 | 81.171 | 89.072 | MANet | 86.959 | 84.444 | 82.688 | 74.923 | 77.255 | 85.420 |

TABLE II
THE EXPERIMENTAL RESULTS ON THE GID DATASET.

| Method | PA | MPA | K | mIoU | FWIoU | F1 |
|---|---|---|---|---|---|---|
| U-Net | 86.378 | 74.532 | 83.357 | 64.516 | 76.822 | 75.532 |
| Deeplab V3 | 89.388 | 80.905 | 87.079 | 71.809 | 81.437 | 81.077 |
| Deeplab V3+ | 90.125 | 81.483 | 87.959 | 72.668 | 82.646 | 81.492 |
| RefineNet | 89.857 | 81.169 | 87.597 | 73.167 | 82.109 | 83.113 |
| PSPNet | 90.573 | 82.211 | 88.485 | 74.797 | 83.255 | 83.761 |
| FastFCN | 90.336 | 83.625 | 88.221 | 74.364 | 82.950 | 83.704 |
| DANet | 90.135 | 83.035 | 87.959 | 74.040 | 82.591 | 83.727 |
| OCRNet | 90.560 | 82.402 | 88.472 | 74.479 | 83.303 | 83.350 |
| MANet | 91.530 | 86.514 | 89.678 | 78.332 | 84.898 | 87.135 |

TABLE III
THE COMPARISON OF PARAMETERS AND COMPUTATIONAL COMPLEXITY, 'M' REPRESENTS MILLION, 'G' DENOTES GILLION (THOUSAND MILLION).

| Method | input shape | Parameters (M) | Complexity (G) |
|---|---|---|---|
| U-Net | | 10.86 | 13.96 |
| Deeplab V3 | | 58.16 | 18.63 |
| Deeplab V3+ | | 59.47 | 23.99 |
| RefineNet | | 109.87 | 60.77 |
| PSPNet | $3 \times 256 \times 256$ | 65.71 | 19.69 |
| FastFCN | | 104.38 | 70.51 |
| DANet | | 66.56 | 20.93 |
| OCRNet | | 70.00 | 39.85 |
| MANet | | 93.65 | 25.73 |

[1], and the fine-resolution Gaofen Image Dataset (GID) [86]. All the results for the ISPRS dataset are tested using ground reference data without eroded boundaries, so the evaluation indices are not as high as reported in certain elements of the literature.

**Potsdam:** The Potsdam dataset contains 38 fine-resolution images of size 6000 × 6000 pixels with a ground sampling distance (GSD) of 5 cm. The dataset provides near-infrared, red, green and blue channels as well as DSM and normalized DSM (NDSM). There are 24 images in the training set and 16 images in the test set. Specifically, we utilize ID: 2_13, 2_14, 3_13, 3_14, 4_13, 4_14, 4_15, 5_13, 5_14, 5_15, 6_13, 6_14, 6_15, 7_13 for testing, ID: 2_10 for validation, and the remaining 22 images, except image named 7_10 with error annotations, for training. Note that we use only the red, green and blue channels in our experiments. For training, we crop the raw images into 256 × 256 patches and augmented them by rotating, resizing, horizontal axis flipping, vertical axis flipping, and adding random noise.

**Vaihingen:** The Vaihingen semantic labeling dataset is composed of 33 images with an average size of 2494 × 2064 pixels and a GSD of 5 cm. The near-infrared, red and green channels together with DSM are provided in the dataset. There are 16 images in the training set and 17 images in the test set. We exploited ID: 2, 4, 6, 8, 10, 12, 14, 16, 20, 22, 24, 27, 29, 31, 33, 35, 38 for testing, ID: 30 for validation, and the remaining 15 images for training. We did not use the DSM in our experiments. The process of the training dataset is identical to that for Potsdam.

**GID:** The GID contains 10 RGB images of size 7200 × 6800 pixels captured by the Gaofen-2 Satellite in China. Each image covers a geographic region of 506 km², and is labeled with 15 classes of land use categories. We partitioned each image separately into non-overlapping patch sets of size 256 × 256 pixels and discarded pixels on the edges that could not be divided by 256. As such, 7280 patches are acquired. We then selected randomly 60% of the patches for training, 20% for validation, and the remaining 20% for testing.

### B. Evaluation Metrics

The performance of MANet on the three datasets is evaluated using the pixel accuracy (PA), mean pixel accuracy (MPA), Kappa coefficient (K), and the mean Intersection over Union (mIoU). The PA tends to be insensitive to minority categories while mIoU is prone to be oversensitive to minority classes in imbalanced datasets. Therefore, we use the F1 score (F1) and the Frequency Weighted Intersection over Union (FWIoU) to further evaluate the performance. The PA, MPA, K, mIoU, F1, and FWIoU metrics are computed as:

$$PA = \frac{\sum_{i=0}^{k} p_{ii}}{\sum_{i=0}^{k} \sum_{j=0}^{k} p_{ij}}, \quad (12)$$

$$MPA = \frac{1}{k+1} \sum_{i=0}^{k} \frac{p_{ii}}{\sum_{j=0}^{k} p_{ij}}, \quad (13)$$

$$K = \frac{p_o - p_e}{1 - p_e}, \quad (14)$$

$$mIoU = \frac{1}{k+1} \sum_{i=0}^{k} \frac{p_{ii}}{\sum_{j=0}^{k} p_{ij} + \sum_{j=0}^{k} p_{ji} - p_{ii}}, \quad (15)$$

$$FWIoU = \frac{1}{\sum_{i=0}^{k} \sum_{j=0}^{k} p_{ij}} \sum_{i=0}^{k} \frac{p_{ii}}{\sum_{j=0}^{k} p_{ij} + \sum_{j=0}^{k} p_{ji} - p_{ii}}, \quad (16)$$

$$F1 = 2 \times \frac{precision \times recall}{precision + recall}. \quad (17)$$

where $p_{ii}$, $p_{ij}$, and $p_{ji}$ represent the number of true positives, false positives, and false negatives, respectively. $k$ is the number of classes.

TABLE IV
THE ABLATION STUDY ON THE ISPRS POTSDAM DATASET.

| Backbone | Attention | PA | MPA | K | mIoU | FWIoU | F1 |
|---|---|---|---|---|---|---|---|
| ResNet-50 | | 87.529 | 86.448 | 83.320 | 77.697 | 78.288 | 87.218 |
| ResNet-50 | √ | 88.603 | 88.135 | 84.732 | 79.416 | 79.914 | 88.327 |
| ResNet-101 | | 88.224 | 88.073 | 84.245 | 78.893 | 79.293 | 87.992 |
| ResNet-101 | √ | 89.082 | 88.904 | 85.400 | 80.142 | 80.681 | 88.778 |
| ResNeXt-50 | | 88.337 | 87.172 | 84.368 | 78.626 | 79.349 | 87.867 |
| ResNeXt-50 | √ | 88.766 | 87.661 | 84.921 | 79.421 | 80.125 | 88.283 |
| ResNeXt-101 | | 88.694 | 88.179 | 84.866 | 79.528 | 80.032 | 88.389 |
| ResNeXt-101 | √ | 89.396 | 88.763 | 85.808 | 80.585 | 81.171 | 89.072 |

TABLE V
THE ABLATION STUDY ON THE GID DATASET.

| Backbone | Attention | PA | MPA | K | mIoU | FWIoU | F1 |
|---|---|---|---|---|---|---|---|
| ResNet-50 | | 90.375 | 83.088 | 88.272 | 73.861 | 83.145 | 83.454 |
| ResNet-50 | √ | 90.526 | 83.938 | 88.440 | 75.271 | 83.320 | 84.793 |
| ResNet-101 | | 90.444 | 81.937 | 88.310 | 74.489 | 83.071 | 84.123 |
| ResNet-101 | √ | 90.921 | 83.771 | 88.890 | 76.719 | 83.830 | 85.831 |
| ResNeXt-50 | | 90.549 | 84.228 | 88.524 | 74.536 | 83.484 | 83.863 |
| ResNeXt-50 | √ | 90.474 | 84.061 | 88.397 | 76.157 | 83.251 | 85.468 |
| ResNeXt-101 | | 91.352 | 85.119 | 89.455 | 77.300 | 84.646 | 86.209 |
| ResNeXt-101 | √ | 91.530 | 86.514 | 89.678 | 78.332 | 84.898 | 87.135 |

## C. Experimental Setting

To evaluate the performance of MANet comprehensively, we consider several benchmark comparators including U-Net [21], DeepLab V3 [22], DeepLab V3+ [23], RefineNet [42], PSPNet [24], FastFCN [27], DANet [40] and OCRNet [47]. All the models are implemented with PyTorch, and the Adam optimizer with a 0.0003 learning rate and 16 batch size [2]. All the experiments are implemented on a single NVIDIA GeForce RTX 2080ti GPU with 11 GB RAM. The cross-entropy loss function is used as a quantitative evaluation coupled with backpropagation to measure the disparity between the achieved segmentation maps and the ground reference.

## D. The Complexity of Kernel Attention

We analyze the efficiency merit of kernel attention over dot-product attention in both memory and computation in this section. Given a feature $X = [x_1, \cdots, x_N] \in \mathbb{R}^{N \times C}$, both the dot-attention and kernel attention will generate the query matrix $Q$, key matrix $K$, and value matrix $V$.

For the dot-attention, to compute the similarity using softmax function, we have to generate the $N \times N$ matrix by multiplying the transposed key matrix $K$ and the value matrix $V$, resulting in $O(D_k N^2)$ time complexity and $O(N^2)$ space complexity. Thus, to compute the similarity between each pair of positions, the dot-attention would occupy at least $O(N^2)$ memory and require $O(D_k N^2)$ computation.

For kernel attention, as the softmax function is substituted for kernel smoothers, we can alter the order of the commutative operation and avoid multiplication between the reshaped key matrix $K$ and query matrix $Q$. Therefore, we can calculate the product between $\text{softplus}(K)^T$ and $V$ first and then multiply

[2] Code will be available at https://github.com/lironui/Multi-Attention-Network.

result and $Q$ with only $O(dN)$ time complexity and $O(dN)$ space complexity.

Dot-attention and kernel attention are compared in terms of resource consumption in Fig. 6. For a $64 \times 64 \times 64$ input, the kernel attention yields a 21-fold saving of memory (69MB to 3MB) and an 89-fold saving of computation (3 GMMACC to 37 MMACC). With increasing input size, the gap widens. For a $64 \times 256 \times 256$ input, the dot-attention requires unreasonable memory (17 GB) and computation (829 GMACC), while the kernel attention utilizes merely 1/340 memory (51MB) and 1/1417 computation (585 MMACC).

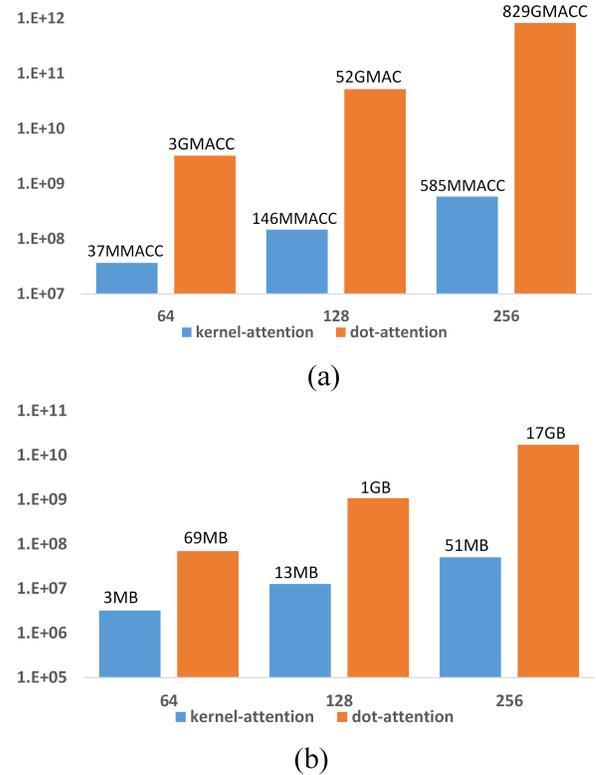

Fig. 6. Computation (a) and memory (b) requirements under different input sizes. The blue and orange bars depict the resource requirements of the kernel attention and dot-attention, respectively. The calculation assumes $D = D_v = 2D_k = 64$. The figure is in log scale.

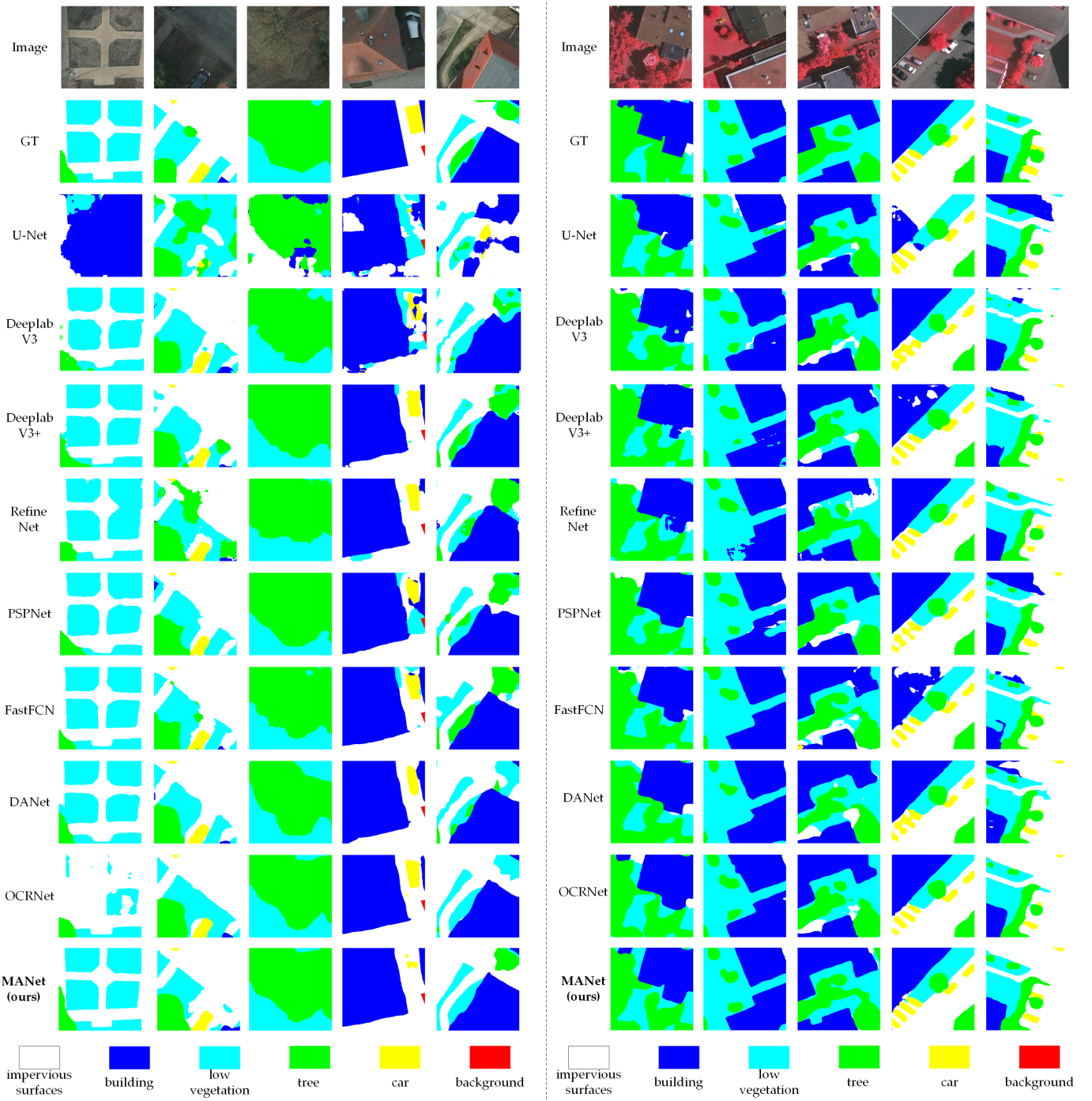

Fig. 7. Qualitative comparisons between different methods on **Left**: the ISPRS Potsdam and **Right**: the Vaihingen dataset.

### E. Results on ISPRS Datasets

The experimental results for the different methods applied to the ISPRS Potsdam and Vaihingen datasets are listed in Table I. The performance of the proposed MANet exceeds that for the other algorithms in all quantitative evaluation indices with a large margin.

As the structure of U-Net is plain and shallow with finite feature extraction capability, the performance of U-Net is inadequate. The ResNet-based frameworks demonstrate greater accuracy, partly attributing to the powerful representation ability of ResNet. Meanwhile, the multi-scale contextual information aggregated using atrous spatial pyramid pooling (DeepLab V3 and DeepLab V3+) [27, 28], pyramid pooling module (PSPNet), [29], joint pyramid upsampling module (FastFCN) or chained residual pooling (RefineNet) [25] contributes to increased accuracy of the final semantic map.

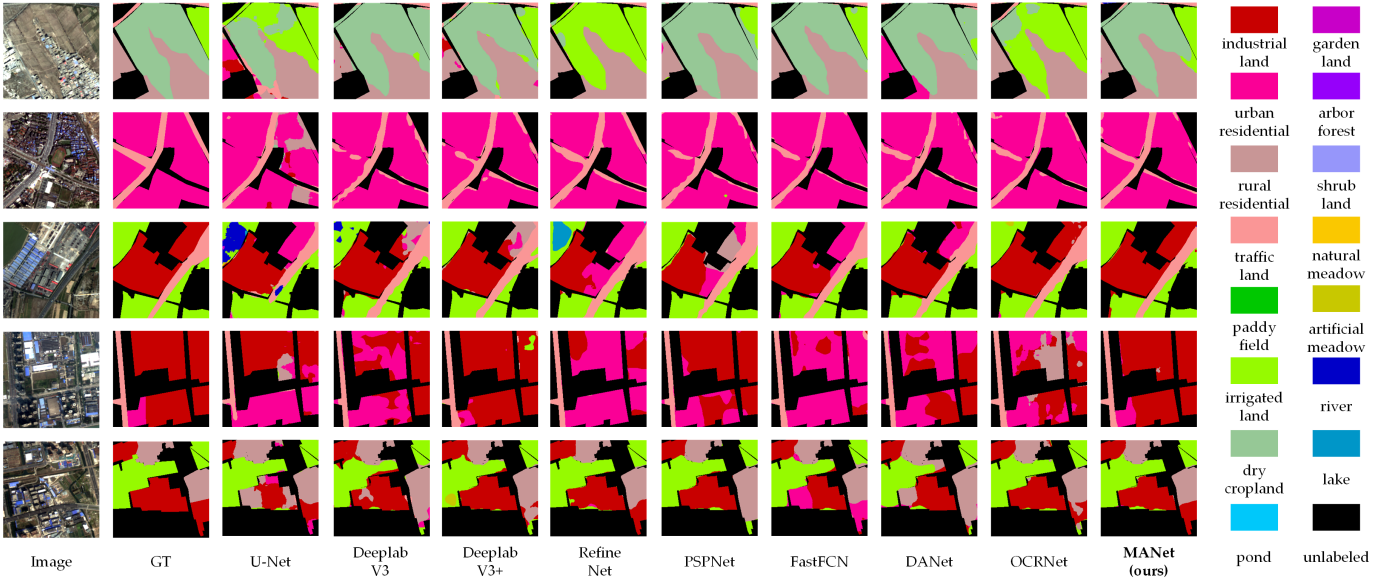

Fig. 8. Qualitative comparisons between different methods on the GID dataset.

Whereas the long-range dependencies are captured by the attention mechanism, DANet [40] and OCRNet [63] perform relatively well.

Instead of capturing contextual information by dilated convolutions, pooling operations or individual attention, we model directly the global dependency of each pair of pixels in the input onto multi-levels. As the demand for memory and computation is reduced dramatically by the proposed kernel attention, it is possible to exploit the multi-level global contextual information, regardless of the size of the input feature maps. Therefore, the design of the proposed MANet, which models long-range dependencies on four scales, is conducive to capturing refined features of the input.

For the ISPRS Potsdam dataset, the observed increase in accuracy is 1.977% in mIoU and 1.226% in F1-score compared with DANet, respectively. For the ISPRS Vaihingen dataset, the increase in accuracy is 2.248%, and 1.568% generated by MANet compared to OCRNet. Some visual results generated by our method and other methods are illustrated in Fig. 7.

### F. Results on GID Datasets

Differing from the ISPRS datasets, which comprise five land cover categories only, 15 classes of land use exist in the GID dataset. Therefore, the segmentation of the GID dataset is more challenging than for the ISPRS datasets. Benefitting from the multi-scale contextual information modeled by the kernel attention mechanisms applied to multi-level layers, the proposed MANet brings 3.535% and 3.374% accuracy increments in the mIoU and F1-score, respectively, compared with PSPNet. Some visual results generated by our method and other methods are illustrated in Fig. 8, which shows that the proposed MANet can capture fine-grained features with high accuracy.

The number of parameters and computational complexity is used to test the parsimoniousness and speed of the MANet framework. As shown in Table III, the complexity of MANet on the GID dataset is reasonable in both memory and computational time: significantly lower than FastFCN and RefineNet, and comparable to other ResNet-based baseline approaches.

### G. Ablation Study

In the proposed MANet, attention blocks and ResNeXt-101 are used to exploit global contextual representations and enhance the capability for feature extraction. To further evaluate the performance of the attention block and backbone, we conduct ablation experiments on the ISPRS Potsdam and GID datasets using different settings than listed in Table IV and Table V. As shown in Table IV and Table V, the utilization of attention blocks and ResNeXt increases the accuracy significantly compared with the baseline methods. For different backbones, the attention mechanism increases the mIoU by 1% on average. ResNeXt-101 and ResNet-101 produce higher accuracies than ResNeXt-50 and ResNet-50 since deeper backbones enable networks to capture more refined features. The ResNeXt-based backbones outperform the ResNet-based backbones, where high accuracy and effectiveness are shown in ResNeXt.

### H. The Effectiveness of Kernel Attention

To further evaluate the effectiveness of the proposed kernel attention, six baselines with kernel attention are applied to GID and their performances are compared. Specifically, the kernel attention is attached to the final decoder of U-Net, the fourth path of RefineNet, the last convolution of FastFCN, and the ResNet-101 backbone of Deeplab v3, Deeplab v3+, PSPNet, and FastFCN. The results are shown in Table VI.

As the representation ability of U-Net is limited, the introduction of kernel attention gives rise to a significant improvement. Concretely, the mIoU is increased by 3.750% and the F1-score by 2.853%. Increases for the other baselines are not as distinct as for U-Net, but still significant: more than 2% improvement in mIoU for DeepLab V3 and DeepLab V3+, and around 1% in mIoU for the others. These results demonstrate that the utilization of kernel attention can capture the global contextual dependencies of feature maps effectively,

TABLE VI
THE EFFECTIVENESS OF KERNEL ATTENTION ON THE GID DATASET.

| Backbone | Attention | PA | MPA | K | mIoU | FWIoU | F1 |
|---|---|---|---|---|---|---|---|
| U-Net | | 86.378 | 74.532 | 83.357 | 64.516 | 76.822 | 75.532 |
| | √ | 88.456 | 77.015 | 85.876 | 68.266 | 79.882 | 78.385 |
| RefineNet | | 89.857 | 81.169 | 87.597 | 73.167 | 82.109 | 83.113 |
| | √ | 90.111 | 84.118 | 87.962 | 75.323 | 82.599 | 85.002 |
| DeepLab V3 | | 89.388 | 80.905 | 87.079 | 71.809 | 81.437 | 81.077 |
| | √ | 90.344 | 82.44 | 88.223 | 73.979 | 82.926 | 82.783 |
| DeepLab V3+ | | 90.125 | 81.483 | 87.959 | 72.668 | 82.646 | 81.492 |
| | √ | 90.804 | 83.498 | 88.787 | 75.328 | 83.738 | 83.054 |
| PSPNet | | 90.573 | 82.211 | 88.485 | 74.797 | 83.255 | 83.761 |
| | √ | 90.739 | 82.957 | 88.688 | 75.496 | 83.542 | 84.424 |
| FastFCN | | 90.336 | 83.625 | 88.221 | 74.364 | 82.950 | 83.704 |
| | √ | 90.820 | 84.026 | 88.814 | 75.182 | 83.754 | 84.118 |

and increase the accuracy of the entire set of baselines.

### I. Discussion on the Attention Mechanism

Selective visual attention endows humans with the ability to orientate towards conspicuous objects over the visual scene in a computationally efficient manner. Thus, the attention mechanism, inspired by the biological mechanism, is intended as a computationally efficient structure with configurable flexibility. By representing the concept of attention via the lens of the kernel [52], we design a kernel attention module with $O(N)$ complexity. The effectiveness and efficiency of the proposed kernel attention is demonstrated consistently across a wide range of quantitative experiments. We envisage the demonstrated resource efficiency will encourage more pervasive and flexible combinations between attention mechanisms and networks.

## V. Conclusion

This paper proposes kernel attention to reduce the complexity of the dot-product attention mechanism into $O(N)$. By integrating kernel attention and ResNeXt-101, we design a Multi-Attention-Network (MANet) comprised of a multi-scale strategy to incorporate semantic information at different levels, together with self-attention modules to aggregate relevant contextual features hierarchically. MANet exploits contextual dependencies over local features producing increased accuracy and computational efficiency. We implement a series of experiments involving the complex task of semantic segmentation of fine-resolution remote sensing images. MANet produces consistently the best classification performance with the highest accuracy. An extensive ablation study is conducted to evaluate the impact of the individual components of the proposed framework. Experimental results on ISPRS and GID datasets demonstrate that the performance of the proposed framework greatly exceeds eight baseline methods across all accuracy indices.

## Acknowledgments

The design of kernel attention in this paper was inspired by the article posted in Jianlin Su's blog, retrieved from https://spaces.ac.cn/archives/7546 (04 Jul 2020).

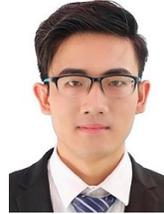

**Rui Li** received a bachelor's degree from the School of Automation Science and Engineering, South China University of Technology, Guangzhou, China in 2019. He is currently pursuing a master's degree with the School of Remote Sensing and Information Engineering, Wuhan University, Wuhan, China.

His research interests include semantic segmentation, hyperspectral image classification, and deep learning.

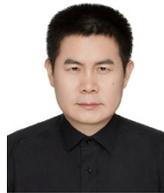

**Shunyi Zheng** received the Post-Doctorate from the State Key Laboratory of Information Engineering in Surveying, Mapping, and Remote Sensing, Wuhan University, Wuhan, China, in 2002. He is currently a Professor at the School of Remote Sensing and Information Engineering, Wuhan University, Wuhan, China. His research interests include remote sensing data processing, digital photogrammetry, and three-dimensional reconstruction. Prof. Zheng received the First Prize for Scientific and Technological Progress in Surveying and Mapping, China, in 2012 and 2019.

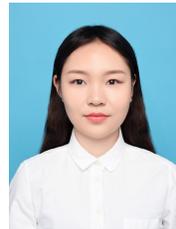

**Chenxi Duan** received a bachelor's degree from the College of Geology Engineering and Geomatics, Chang'an University, Xi'an, China in 2019. She is currently pursuing a master's degree with the State Key Laboratory of Information Engineering in Surveying, Mapping, and Remote Sensing, Wuhan University, Wuhan, China.

Her research interests include cloud removal, numerical optimization, and machine learning.

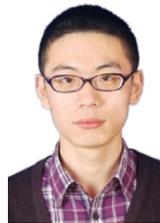

**Ce Zhang** received Ph.D. Degree in Geography from Lancaster Environment Centre, Lancaster University, U.K. in 2018. He was the recipient of a prestigious European Union (EU) Erasmus Mundus Scholarship for a European Joint MSc programme between the University of Twente (The Netherlands) and the University of Southampton (U.K.). Dr. Zhang is currently a Lecturer in Geospatial Data Science at the Centre of Excellence in Environmental Data Science (CEEDS), jointly venture between Lancaster University and UK Centre for Ecology & Hydrology (UKCEH). His major research interests include geospatial artificial intelligence, machine learning, deep learning, and remotely sensed image analysis.

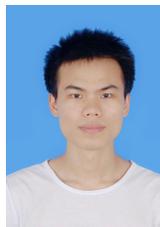

**Jianlin Su** received the master's degree from the School of Mathematics, Sun Yat-sen University, Guangzhou, China. He is currently the senior researcher in the Shenzhen Zhuiyi Technology Co., Ltd

His research interests are focused on the generative model, including the language model and Seq2Seq in natural language processing, and the GAN, VAE, and flow in the computer vision. Besides, he is also interested in the basic theory of machine learning. His homepage is http://jianlin.su.

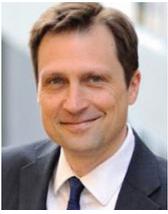

**Peter M. Atkinson** received the MBA degree from the University of Southampton, Southampton, U.K. in 2012, and the Ph.D. degree from the University of Sheffield, Sheffield, U.K. (NERC CASE Award with Rothamsted Experimental Station), in 1990.

He is currently the Dean of the Faculty of Science and Technology with Lancaster University, U.K. He was previously a Professor of geography with the University Southampton, where he is currently a Visiting Professor. He is also a Visiting Professor with Queen's University Belfast, U.K., and with the Chinese Academy of Sciences, Beijing, China. He has authored more than 270 peer-reviewed articles in international scientific journals and around 50 refereed book chapters. He has also edited nine journal special issues and eight books. His research interests include remote sensing, geographical information science, and spatial (and space-time) statistics applied to a range of environmental science and socio-economic problems.

Prof. Atkinson is an Editor-in-Chief of Science of Remote Sensing, a sister journal of Remote Sensing of Environment. He is also an Associate Editor for the Computers and Geosciences and sits on the editorial boards of several further journals including Geographical Analysis, Spatial Statistics, the International Journal of Applied Earth Observation and Geoinformation, and Environmental Informatics. He sits on various international scientific committees. He previously held the Belle van Zuylen Chair with Utrecht University, The Netherlands and is recipient of the Peter Burrough Award of the International Spatial Accuracy Research Association.